\documentclass[prd,floats,aps,nofootinbib,amssymb,11pt]{revtex4}
\usepackage{amsmath}
\usepackage{amsthm}
\usepackage{float}
\usepackage{graphicx}
\usepackage{bm}
\usepackage{epstopdf}
\usepackage{color}
\usepackage{slashed}
\usepackage{hyperref}

\newcommand{\beq}{\begin{equation}}
\newcommand{\eeq}{\end{equation}}
\newcommand{\bea}{\begin{eqnarray}}
\newcommand{\eea}{\end{eqnarray}}
\newcommand{\ba}{\begin{array}}
\newcommand{\ea}{\end{array}}

\def\m1{M_1}
\def\m2{M_2}
\def\m3{M_3}

\def\ch10{\tilde \chi^0_1}

\def\gev{\,{\rm GeV}}

\def\to{\rightarrow}

\newcommand{\lsim}{\mathrel{\mathop{\kern 0pt \rlap
  {\raise.2ex\hbox{$<$}}}
  \lower.9ex\hbox{\kern-.190em $\sim$}}}
\newcommand{\gsim}{\mathrel{\mathop{\kern 0pt \rlap
  {\raise.2ex\hbox{$>$}}}
  \lower.9ex\hbox{\kern-.190em $\sim$}}}

\newcommand{\ud}{\mathrm{d}}
\newcommand{\ee}{{$e^{-}e^{+}$}}
\newcommand{\fb}{{\,{\rm fb}}}

\begin{document}

\title{Double interference effects in Higgs precision tests in the $e^-e^+\to \nu\bar \nu h$ process}

\author{Yang Zhang\footnote{E-mail: zhang-yang13@mails.tsinghua.edu.cn}}
\affiliation{Department of Physics, Tsinghua University, Beijing, 100084, China}
\affiliation{Center for High Energy Physics, Tsinghua University, Beijing, 100084, China}

\begin{abstract}
Higgs precision program at future lepton collider aims at (sub) percent level precision measurement of the Higgs properties, shedding light to new physics through the Higgs lamppost. Amongst many exclusive Higgs channels that can be measured precisely, the $WW$-fusion to Higgs with subsequent decays into $b\bar b$ final state are of particular importance. This channel provide leading constrains on Higgs total width in the $\kappa$-framework and greatly improves the constraints in the EFT framework as a distinct production mode other than the Higgsstrahlung process. We argue in this paper that, there are two interference effects both affects the physical information one can extract from the precision measurements. One takes place at quantum level from the interference between the two amplitudes that amounts to $-10\%$ of the $WW$-fusion signal strength, failing to take into account which will result in a $4$-$5\sigma$ discrepancy between theory and measurement. The other takes place at the classical level from the global fitting procedure where the $ZH$ process is the dominant background with its cross section around six times larger than the $WW$-fusion signal. Despite that $ZH$ process can be measured to great precision at future lepton colliders, failing take this interplay in the coupling extraction will result in a 100\% too aggressive constraints on $\kappa_W$, the Higgs coupling to $W$-boson pairs. This effect will also be important for lepton colliders running at slightly higher energies where the phase space overlap are still sizable between the two processes.
\end{abstract}

\maketitle

\section{Introduction}

In the summer of 2012, LHC established the existence of the Higgs boson, which completed the Standard Model (SM) \cite{Aad:2012tfa,Chatrchyan:2012xdj}. However, instead of being an end, this ``God particle" marks the start of a new journey. In spite of the great triumph of the SM, it is not a final theory of particle physics. For example, there is compelling evidence, both astronomically and cosmologically, for the existence of dark matter, while there is no particle candidate for dark matter in the SM. 
Moreover, the discovery of the Higgs boson has also left us with many open questions. As a fundamental scalar particle, Higgs mass is quadratically sensitive to the new physics scale, which implies this new scale should not be too far above the Electroweak (EW) scale. To extend the particle physics theory and settle these problems, different kinds of models have been developed, such as Supersymmetry (SUSY) and composite Higgs model. Since Higgs physics plays an essential rule in constraining and evaluating these models, a detailed understanding of Higgs sector is obviously of great significance to the exploration of new physics. Therefore a precision measurement program of Higgs physics will be the major focus of high energy physics in the coming decades.

Due to the contributions of the LHC experiments, some of the Higgs properties, such as the Higgs couplings to $W/Z$ bosons and gluons have already been measured with a precision level at about $\mathcal{O}(15\%)$. This will be further improved with more data from LHC Run 2 and in future upgrade program. However, because of the composite nature of the proton, the LHC can only provide a fuzzy picture of the Higgs. To achieve sub-percent level of precision, an electron-positron collider will be required to operate as a Higgs factory, such as the planned CEPC \cite{cepccdr} and FCC-ee \cite{Gomez-Ceballos:2013zzn}. Running at a center-of-mass (c.m.) energy of about 250 GeV, such a Higgs factory will reach its peak value of the Higgs production cross section through the so called Higgsstrahlung process (or $ZH$ process) $e^+e^-\to ZH$. Since the c.m. energy at an $e^+e^-$ collider is precisely measurable, the Higgs boson mass can be reconstructed through the recoil mass method and Higgs candidate events can be identified. In the selected $e^+e^-\to ZH$ events, branching ratios of different Higgs decay modes can then be measured by seperating different decay final states.

While, the total Higgs boson decay width is not that straightforward to get. The total decay width of a 125 GeV SM Higgs boson $\Gamma_H$ is 4 MeV \cite{Patrignani:2016xqp}, which is too small to be measured directly. However, it can be indirectly derived from some of its decay channels using decay width and other relevant accessible quantities. One proper way is through the Higgs decaying to $ZZ^\ast$ \cite{cepccdr}:
\begin{equation}
\Gamma_H = \dfrac{\Gamma(H\to ZZ^{\ast})}{{\rm BR}(H\to ZZ^{\ast})}\propto
  \dfrac{\sigma(ZH)}{{\rm BR}(H\to ZZ^{\ast})}
\end{equation}
Here the partial decay width $\Gamma(H\rightarrow ZZ^\ast)$ depends on the Higgs-$Z$ boson coupling $g(HZZ)$ which is then substituted equivalently by the cross section of the Higgsstrahlung process $\sigma(ZH)$. In this case, the performance of $H\rightarrow ZZ^\ast$ measurement will determine the precision of $\Gamma_H$ value.

Alternatively, we can turn to another decay mode of Higgs to a pair of $b$-quarks $H\to b\bar{b}$:
\begin{equation}\label{width2}
\Gamma_H = \dfrac{\Gamma(H\to b\bar{b})}{{\rm BR}(H\to b\bar{b})}.
\end{equation}
The partial width $\Gamma(H\to b\bar{b})$ can be obtained from the $WW$ fusion process with the Higgs decaying to a pair of $b$-jets $e^+e^-\to \nu\bar{\nu}H\to \nu\bar{\nu}b\bar{b}$:
\begin{align}
\sigma(\nu\bar{\nu}H\to \nu\bar{\nu}b\bar{b})& \propto \Gamma(H\to WW^{\ast})\cdot {\rm BR}(H\to b\bar{b})\\
&\propto {\rm BR}(H\to WW^{\ast})\cdot\Gamma(H\to b\bar{b})
\end{align}
Replacing $\Gamma(H\to b\bar{b})$ in Eq.~(\ref{width2}) with the two measured quantities $\sigma(\nu\bar{\nu}H\to \nu\bar{\nu}b\bar{b})$ and BR$(H\to WW^{\ast})$, the total width $\Gamma_H$ is written as:
\begin{equation}
\Gamma_H\propto \dfrac{\sigma(\nu\bar{\nu}H\to \nu\bar{\nu}b\bar{b})}{{\rm BR}(H\to b\bar{b})\cdot{\rm BR}(H\to WW^{\ast})}
\end{equation}
Here the branching ratios BR$(H\to b\bar{b})$ and BR$(H\to WW^{\ast})$ are extracted from the inclusive cross section $\sigma(ZH)$ and the cross sections of individual Higgs decay mode $\sigma(ZH)\times{\rm BR}$, as mentioned before. The precision of the total width achieved with this method is limited by the small cross section $\sigma(\nu\bar{\nu}H\to \nu\bar{\nu}b\bar{b})$, which is the main concern of this paper. At an $e^+e^-$ collider with c.m. energy about 250 GeV, the $WW$ fusion process is also one of the leading production processes for a 125 GeV SM Higgs boson, together with the Higgsstrahlung process. And with $Z\to \nu\bar{\nu}$, the $ZH$ channel interferes with $WW$ fusion channel, contributing an irreducible background to the $\nu\bar\nu b\bar b$ final state. In the following content, we first calculate the cross sections of the $WW$ fusion channel and the interference effect both analytically and numerically in Section.~\ref{sec1}. Then, some important differential distributions of both the signal and the background channels are presented in Section.~\ref{sec3}. Finally, we analyze the impact of this interference effect on the Higgs physics measurement at $e^+ e^-$ colliders.

\section{The interference effects}\label{sec1}

\subsection{Analytic understanding}

\begin{figure}
  \centering
  \includegraphics[scale=1,clip]{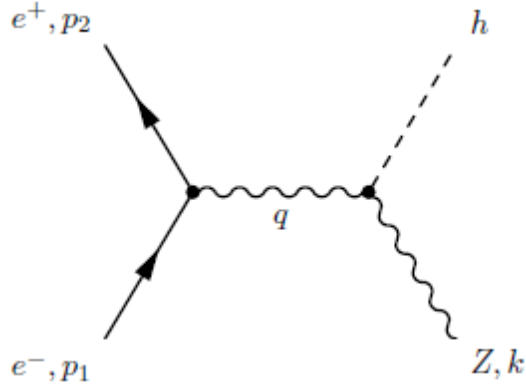}
  \caption{
  Feynman diagram contributing to the amplitude of $e^+e^- \to ZH$ process.
  }
\label{fig:zh}
\end{figure}
Fig.~\ref{fig:zh} shows the Feynman diagram for the Higgsstrahlung process, which is one with $s$-channel $Z$ boson exchange. Since left-handed and right-handed electrons have different gauge couplings, evaluating the diagram gives amplitudes respectively for $e^{-}_L e^{+}_R$ and $e^{-}_R e^{+}_L$ initial states:
\begin{align}
i\mathcal{M}(e_L^- e_R^+)&= \bar{v}_L(p_2)\frac{ie\gamma_\mu}{c_w s_w}
(-\frac{1}{2}+s_w^2)u_L(p_1)\cdot\frac{-ig^{\mu\nu}}{q^2-m_z^2}\cdot
2i\frac{m_z^2}{v}g_{\nu\rho}\cdot\epsilon^{\ast\rho}(k)\\
&=\frac{ie}{c_w s_w}(-\frac{1}{2}+s_w^2)\frac{1}{q^2-m_z^2}\frac{2m_z^2}{v}
\cdot \bar{v}_L(p_2)\gamma^\mu u_L(p_1)\epsilon^\ast_\mu(k)\\
i\mathcal{M}(e_R^- e_L^+)&= \bar{v}_R(p_2)\frac{ie\gamma_\mu}{c_w s_w}s_w^2 u_R(p_1)
\cdot\frac{-ig^{\mu\nu}}{q^2-m_z^2}\cdot
2i\frac{m_z^2}{v}g_{\nu\rho}\cdot\epsilon^{\ast\rho}(k)\\
&=\frac{ie}{c_w s_w}s_w^2\frac{1}{q^2-m_z^2}\frac{2m_z^2}{v}
\cdot \bar{v}_R(p_2)\gamma^\mu u_R(p_1)\epsilon^\ast_\mu(k)
\end{align}
where $c_w$ and $s_w$ are respectively $c_w\equiv \cos\theta_W$ and $s_w\equiv \sin\theta_W$. The squared amplitudes are
\begin{align}
|\mathcal{M}(e_L^- e_R^+)|^2 &= \frac{e^2}{c_w^2 s_w^2}(-\frac{1}{2}+s_w^2)^2
\left|\frac{1}{q^2-m_z^2}\right|^2\frac{4m_z^4}{v^2}\epsilon^\ast_\mu \epsilon_\nu\cdot
\bar{v}_L(p_2)\gamma^\mu u_L(p_1)\bar{u}_L(p_1)\gamma^\nu v_L(p_2)\\
&=\frac{e^2}{c_w^2 s_w^2}(-\frac{1}{2}+s_w^2)^2\left|\frac{1}{q^2-m_z^2}\right|^2
\frac{4m_z^4}{v^2}\epsilon^\ast_\mu \epsilon_\nu\cdot
2(p_2^\mu p_1^\nu+p_2^\nu p_1^\mu-p_1\cdot p_2g^{\mu\nu}+i\epsilon^{\alpha\mu\beta\nu}p_{2\alpha}p_{1\beta})\\
|\mathcal{M}(e_R^- e_L^+)|^2 &= \frac{e^2}{c_w^2 s_w^2}s_w^4\left|\frac{1}{q^2-m_z^2}\right|^2
\frac{4m_z^4}{v^2}\epsilon^\ast_\mu \epsilon_\nu\cdot
2(p_2^\mu p_1^\nu+p_2^\nu p_1^\mu-p_1\cdot p_2g^{\mu\nu}-i\epsilon^{\alpha\mu\beta\nu}p_{2\alpha}p_{1\beta})
\end{align}
Summing over final states polarization and averaging over initial states polarization, the unpolarized squared amplitude is,
\begin{align}
\frac{1}{4}\sum|\mathcal{M}|^2 &= \frac{1}{4}\frac{e^2}{c_w^2 s_w^2}(-\frac{1}{2}+s_w^2)^2
\left|\frac{1}{q^2-m_z^2}\right|^2\frac{4m_z^4}{v^2}(-g^{\mu\nu}+\frac{k^\mu k^{\nu}}{m_z^2})\cdot
2(p_2^\mu p_1^\nu+p_2^\nu p_1^\mu-p_1\cdot p_2g^{\mu\nu}+i\epsilon^{\alpha\mu\beta\nu}p_{2\alpha}p_{1\beta})\\
&+ \frac{1}{4}\frac{e^2}{c_w^2 s_w^2}s_w^4\left|\frac{1}{q^2-m_z^2}\right|^2\frac{4m_z^4}{v^2}
(-g^{\mu\nu}+\frac{k^\mu k^{\nu}}{m_z^2})\cdot
2(p_2^\mu p_1^\nu+p_2^\nu p_1^\mu-p_1\cdot p_2g^{\mu\nu}-i\epsilon^{\alpha\mu\beta\nu}p_{2\alpha}p_{1\beta})\\
&=\frac{1}{4}(2s_w^4-s_w^2+\frac{1}{4})\frac{e^2}{c_w^2 s_w^2}
\left|\frac{1}{q^2-m_z^2}\right|^2\frac{4m_z^4}{v^2}[2p_1\cdot p_2+\frac{4}{m_z^2}(k\cdot p_1)(k\cdot p_2)]\\
&=\frac{1}{4}(2s_w^4-s_w^2+\frac{1}{4})\frac{e^2}{c_w^2 s_w^2}
\left|\frac{1}{s-m_z^2}\right|^2\frac{4m_z^4 s}{v^2}[2+\frac{|\vec{k}|^2}{m_z^2}(1-\cos^2\theta)]
\end{align}
where $|\vec{k}|=\dfrac{1}{2\sqrt{s}}\sqrt{s^2-2(m_1^2+m_2^2)s+(m_1^2-m_2^2)^2}$ is the final state momentum in c.m. frame, and $\sqrt{s}$ is the c.m. energy. Then the cross section is
\begin{align}
\frac{\ud\sigma}{\ud\Omega}&=\frac{1}{2s}\frac{|\vec{k}|}{16\pi^2\sqrt{s}}
\cdot\frac{1}{4}\sum|\mathcal{M}|^2 \\
\sigma&= \frac{|\vec{k}|}{8\pi \sqrt{s}}(2s_w^4-s_w^2+\frac{1}{4})\frac{\alpha}{c_w^2 s_w^2}
\left|\frac{1}{s-m_z^2}\right|^2\frac{m_z^4}{v^2}\cdot\int\ud\Omega
(2+\frac{|\vec{k}|^2}{m_z^2}-\frac{|\vec{k}|^2}{m_z^2}\cos^2\theta)\\
&=\frac{|\vec{k}|}{8\pi \sqrt{s}}(2s_w^4-s_w^2+\frac{1}{4})\frac{\alpha}{c_w^2 s_w^2}
\left|\frac{1}{s-m_z^2}\right|^2\frac{m_z^4}{v^2}
(8\pi+\frac{8\pi}{3}\frac{|\vec{k}|^2}{m_z^2})\\
&=\frac{\pi\alpha^2}{24}\bigg(\frac{2|\vec{k}|}{\sqrt{s}}\bigg)
\frac{|\vec{k}|^2+3m_z^2}{(s-m_z^2)^2}\frac{1-4s_w^2+8s_w^4}{s_w^4(1-s_w^2)^2}
\label{eq:ZH}
\end{align}
where $\alpha$ is the fine structure constant. At $\sqrt{s}=250$ GeV, the tree level cross section of the Higgsstrahlung process is 241.556 fb, which agrees with the numerical result computed by CalcHEP \cite{Belyaev:2012qa}. If we take the branching ratio of the $Z$ boson invisible decay to be 20\%, then $\sigma(ZH,Z\rightarrow \nu\bar{\nu})=241.556\fb\times 20\%=48.311\fb$.

\begin{figure}
  \centering
  \includegraphics[scale=0.8,clip]{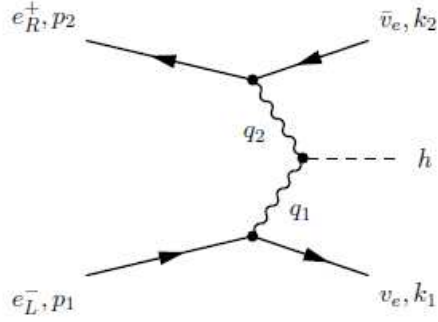}
  \caption{
  Feynman diagram contributing to the amplitude of $WW$ fusion process.
  }
\label{fig:ww}
\end{figure}

The Feynman diagram for $WW$ fusion process is shown in Fig.~\ref{fig:ww}.
There is only one possible polarization configuration for the initial states of this process, $e^{-}_L e^{+}_R$. The polarization amplitude is
\begin{align}
i\mathcal{M}(e_L^- e_R^+)&= \bar{u}_L(k_1)\frac{ig}{\sqrt{2}}\gamma_\mu u_L(p_1)
\cdot\frac{-ig^{\mu\nu}}{q_1^2-m_w^2}\cdot 2i\frac{m_w^2}{v}g_{\nu\rho}\cdot
\frac{-ig^{\rho\lambda}}{q_2^2-m_w^2}\bar{v}_L(p_2)\frac{ig}{\sqrt{2}}\gamma_\lambda v_L(k_2)\\
&=\frac{i4\pi\alpha m_w^2}{s_w^2 v}\frac{1}{q_1^2-m_w^2}\frac{1}{q_2^2-m_w^2}
\bar{u}_L(k_1)\gamma^\mu u_L(p_1)\bar{v}_L(p_2)\gamma_\mu v_L(k_2)\label{eq:wwamp}
\end{align}
Squared amplitude for unpolarized beam is
\begin{align}
\frac{1}{4}\sum|\mathcal{M}|^2 &= \frac{1}{4}\frac{16\pi^2\alpha^2 m_w^4}{s_w^4 v^2}
\left|\frac{1}{q_1^2-m_w^2}\right|^2\left|\frac{1}{q_2^2-m_w^2}\right|^2\\
&\cdot \bar{u}_L(k_1)\gamma^\mu u_L(p_1)\bar{u}_L(p_1)\gamma^\nu u_L(k_1)
\bar{v}_L(p_2)\gamma_\mu v_L(k_2)\bar{v}_L(k_2)\gamma_\nu v_L(p_2)\\
&= \frac{1}{4}\frac{16\pi^2\alpha^2 m_w^4}{s_w^4 v^2}
\left|\frac{1}{2p_1\cdot k_1+m_w^2}\right|^2\left|\frac{1}{2p_2\cdot k_2+m_w^2}\right|^2
\cdot 16(p_1\cdot k_2)(p_2\cdot k_1)\\
&= \frac{(4\pi\alpha)^3 m_w^2}{s_w^6}\left|\frac{1}{2p_1\cdot k_1+m_w^2}\right|^2
\left|\frac{1}{2p_2\cdot k_2+m_w^2}\right|^2\cdot(p_1\cdot k_2)(p_2\cdot k_1)
\end{align}
It is not straightforward to quote analytic results for the total cross section. However, integrating over the 3-body phase space numerically, we can get the result, which is $\sigma(WW~\mathrm{fusion})=7.950\fb$, at $\sqrt{s}=250$ GeV.

\begin{figure}
  \centering
  \includegraphics[scale=0.8,clip]{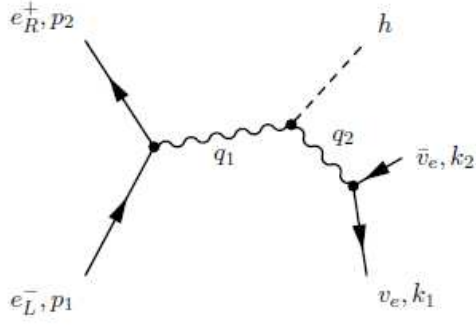}
  \caption{
  Feynman diagram contributing to the amplitude of $e^+ e^- \to Z H \to \nu\bar{\nu}H$ process.
  }
\label{fig:zh_vvh}
\end{figure}

Fig.~\ref{fig:zh_vvh} gives Feynman diagram of another contribution to \ee $\to \nu\bar{\nu}H$ process, which is $e^-e^+\to ZH \to \nu\bar{\nu}H$.
With $Z$ boson off-shell and $e^{-}_L e^{+}_R$ as initial states, this diagram will interfere with $WW$ fusion term. The amplitude for the $ZH$ channel is
\begin{align}
i\mathcal{M}(e_L^- e_R^+)&= \bar{v}_L(p_2)\frac{ie\gamma_\mu}{c_w s_w}
(-\frac{1}{2}+s_w^2)u_L(p_1)\cdot\frac{-ig^{\mu\nu}}{q_1^2-m_z^2}\cdot
2i\frac{m_z^2}{v}g_{\nu\rho}\cdot\frac{-ig^{\rho\lambda}}{q_2^2-m_z^2+im_z\Gamma_Z}
\cdot\bar{u}_L(k_1)\frac{ie\gamma_\lambda}{2c_w s_w}v_L(k_2)\\
&=\frac{i4\pi\alpha}{c_w^2 s_w^2}(-\frac{1}{2}+s_w^2)\frac{m_z^2}{v}\frac{1}{q_1^2-m_z^2}
\frac{1}{q_2^2-m_z^2+im_z\Gamma_Z}\cdot \bar{v}_L(p_2)\gamma^\mu u_L(p_1)
\bar{u}_L(k_1)\gamma_\mu v_L(k_2)
\end{align}
Combining this amplitude and the $WW$ fusion amplitude in Eq.~(\ref{eq:wwamp}), we get the interference effect for unpolarized initial states:
\begin{align}
\frac{1}{4}\sum|\mathcal{M}|^2_{\mathrm{intf}}=& -\frac{1}{4}
(\mathcal{M}_1^\ast\mathcal{M}_2+\mathcal{M}_2^\ast\mathcal{M}_1)\\
=& -\frac{1}{4}\frac{16\pi^2\alpha^2 m_w^2}{c_w^2 s_w^4 v}(-\frac{1}{2}+s_w^2)\frac{m_z^2}{v}
\frac{1}{s-m_z^2}\frac{1}{2p_1\cdot k_1+m_w^2}\frac{1}{2p_2\cdot k_2+m_w^2}\cdot\\
& \frac{4k_1\cdot k_2-2m_z^2}{(2k_1\cdot k_2-m_z^2)^2+m_z^2\Gamma_Z^2}\times
(-16)(p_1\cdot k_2)(p_2\cdot k_1)\\
=& -\frac{(4\pi\alpha)^3 m_w^2}{2s_w^6(1-s_w^2)^2}\bigg(\frac{1-2s_w^2}{s-m_z^2}\bigg)
\frac{1}{2p_1\cdot k_1+m_w^2}\frac{1}{2p_2\cdot k_2+m_w^2}\cdot\\
& \frac{4k_1\cdot k_2-2m_z^2}{(2k_1\cdot k_2-m_z^2)^2+m_z^2\Gamma_Z^2}\times
(p_1\cdot k_2)(p_2\cdot k_1)
\end{align}
Integrating numerically, the contribution from the interference term at $\sqrt{s}=250$ GeV is $\sigma_{\rm intf}=-0.327\fb$. Then the interference effect at $\sqrt{s}=250$ GeV derived from these data is
\begin{equation}
\sigma_{\rm intf}/\sigma(WW~\mathrm{fusion})=\frac{-0.327\fb}{7.950\fb}\approx -4.1\%
\end{equation}

The numerical values of the parameters used to compute those results are summarized below.
\begin{center}
\begin{tabular}{|c|c|c|c|c|c|}
  \hline
  $m_Z$[GeV] & $\Gamma_Z$[GeV] & $G_F$[GeV$^{-2}$]& $\alpha_{em}(m_Z^2)^{-1}$ & $\sin^2\theta_W$ & $m_h$[GeV] \\
  \hline
  91.1876 $\pm$ 0.0021 & 2.441404 & $1.1663787(6)\times10^{-5}$ & 127.940 $\pm$ 0.014 & 0.23124(6) & 125.09 $\pm$ 0.24 \\
  \hline
\end{tabular}
\end{center}

\subsection{Phenomenological analysis}

Following the calculation above, we further perform a numerical study of this interference effect. Letting Higgs boson decay to $b\bar{b}$, then the signal event is two $b$-tagged jets plus missing energy $\nu\bar{\nu}$. We generate the signal and background events for an $\sqrt{s}=250$ GeV $e^-e^+$ collider with the CalcHEP at parton level, and impose the detector acceptance and resolution with our own code. Since the proposed future lepton colliders in general have similar detector performance, here we take performance of the CEPC detector as benchmark for our numerical analysis \cite{cepccdr}.

The $b$-tagging efficiency is conservatively chosen to be 80\%. To achieve this identification efficiency, $b$ jets are required to have $|\cos\theta_j|<0.98$ (or equivalently $|\eta_j|<2.3$) and $|E_j > 10|$ GeV. Since these two $b$ jets are Higgs decay product, the invariant mass is required to satisfy $|m_{jj}-m_h|<5$ GeV.
We mimic the detector resolution effect by adding Gaussian smearing effects on the four-momentum of final state particles. The energy resolution of jets is affected by the hadron calorimeter, and performs approximately as
\begin{equation}
\frac{\delta E}{E}=\frac{0.3}{\sqrt{E/\mathrm{GeV}}}\oplus 0.02.
\end{equation}

Cross sections of the involved different channels before and after imposing the above detector simulation are
\begin{center}
\begin{tabular}{lcc}
  \hline\hline
  Process & no cuts (fb) & \quad after cuts (fb) \quad\\
  \hline
  $e^+ e^-\rightarrow\nu\bar{\nu}H\rightarrow \nu\bar{\nu}b\bar{b}$ & 32.69 & 11.07\\
  $e^+ e^-\rightarrow ZH \rightarrow \nu\bar{\nu}b\bar{b}$ & 28.26 & 9.716\\
  $WW \mathrm{fusion}, H\rightarrow b\bar{b}$ & 4.618 & 1.448\\
  \hline
\end{tabular}
\end{center}
Here the branching ratio of Higgs decaying to $b\bar{b}$ is BR$(H\rightarrow b\bar{b})=58.09\%$ \cite{deFlorian:2016spz}.

The interference effect corresponding to results before cuts is
\begin{align}
\sigma_{\rm intf}/\sigma_{WW}=& \frac{\sigma_{\rm tot}-\sigma_{ZH}-\sigma_{WW}}{\sigma_{WW}}\nonumber  \\
=& \frac{(32.69 - 28.26 - 4.618)\fb}{4.618\fb}=\frac{-0.188\fb}{4.618\fb}\approx -4.1\%
\end{align}
where $\sigma_{\rm tot}$ is the total cross section for the process $e^+ e^-\rightarrow \nu\bar{\nu}H\rightarrow \nu\bar{\nu}b\bar{b}$, and $\sigma_{ZH}$, $\sigma_{WW}$ and $\sigma_{\rm intf}$ are respectively the three components Higgsstrahlung, $WW$ fusion and the interference term. The interference effect corresponding to cross sections after cuts is
\begin{align}
\sigma_{\rm intf}/\sigma_{WW}=& \frac{\sigma_{\rm tot}-\sigma_{ZH}-\sigma_{WW}}{\sigma_{WW}}\nonumber  \\
=& \frac{(11.07 - 9.716 - 1.448)\fb}{1.448\fb}=\frac{-0.094\fb}{1.448\fb}\approx -6.5\%
\end{align}
The $b$-jet energy resolution, as the Gaussian smearing effect in our simulation, has a negative influence on the result. The precision would improve with better $b$-jet energy resolution.

\section{differential distributions}\label{sec3}
\begin{figure}
  \centering
  \begin{tabular}{cc}
  \includegraphics[width=0.5\textwidth,clip]{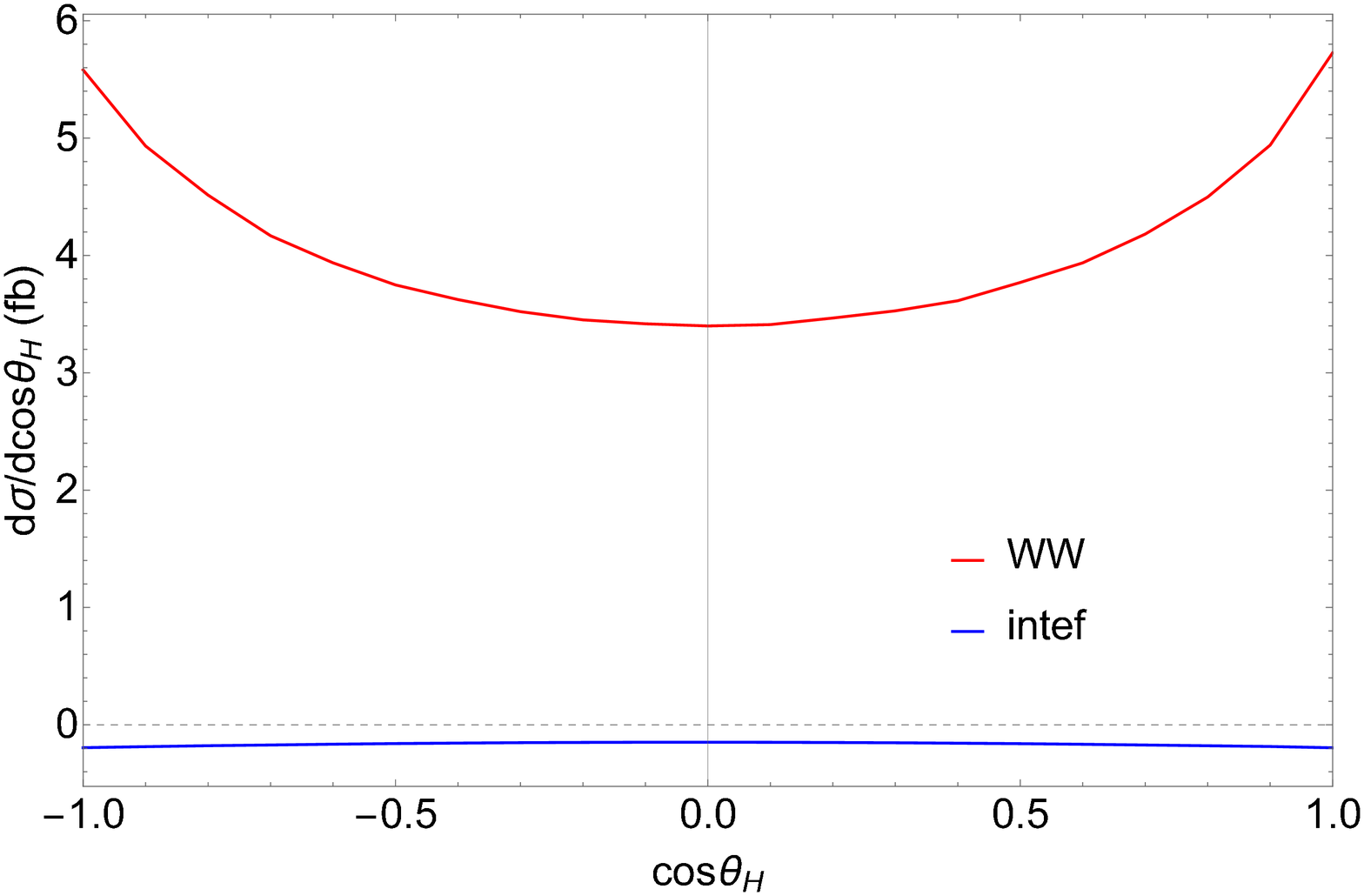}&
  \includegraphics[width=0.5\textwidth,clip]{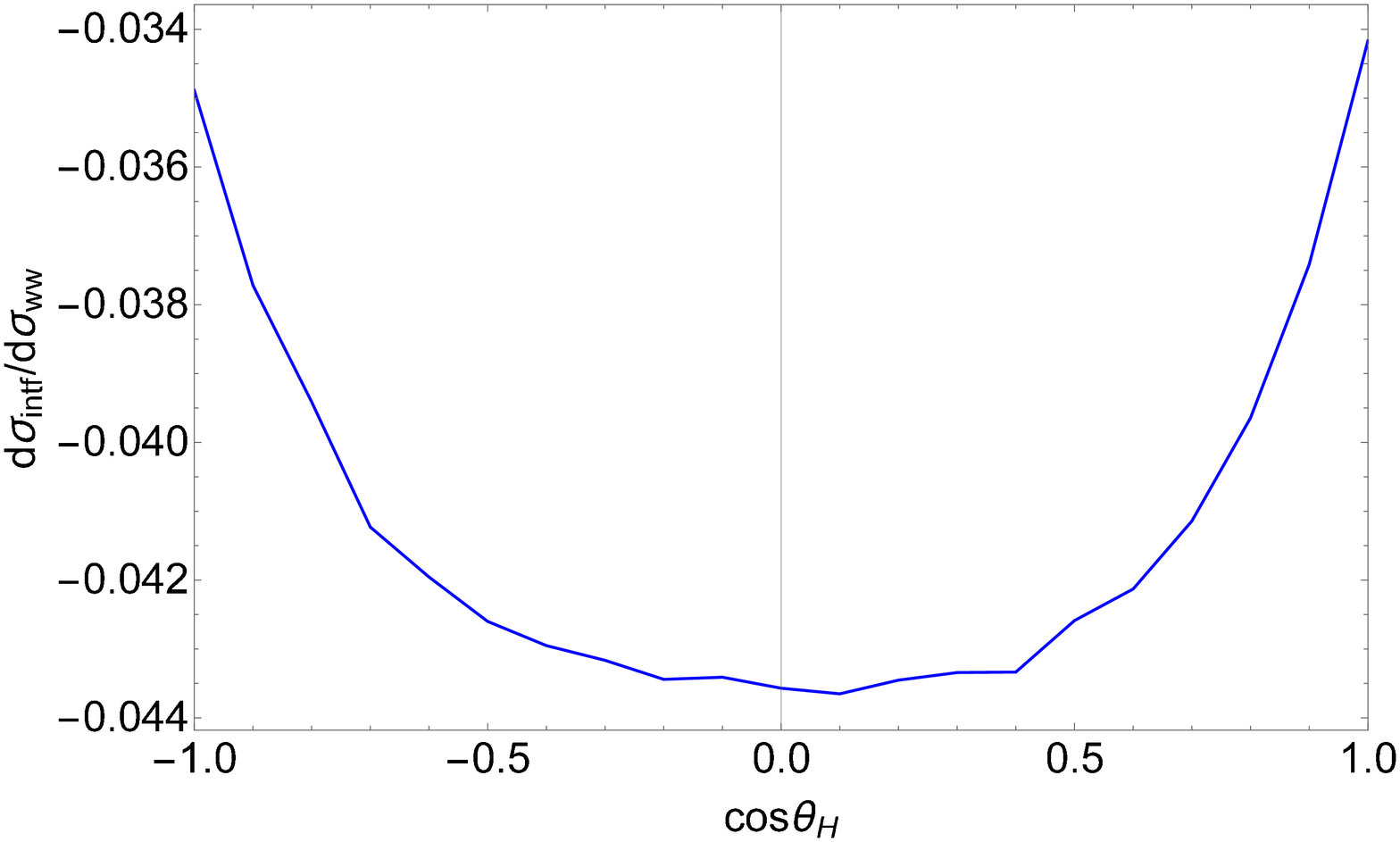}\\
  (a)&(b)\\
  \end{tabular}

  \caption{
    (a) Angular distributions in $\cos\theta_H$ of $WW$ fusion term (red line) and the interference term (blue line);
    (b) the ratio between the angular distributions of the interference term and that of $WW$ fusion term $\ud\sigma_{\rm intf}/\ud\sigma_{WW}$ for different $\cos\theta_H$.
    }
  \label{fig:dsigmacosh}
\end{figure}
\begin{figure}
  \centering
  \begin{tabular}{cc}
  \includegraphics[width=0.5\textwidth,clip]{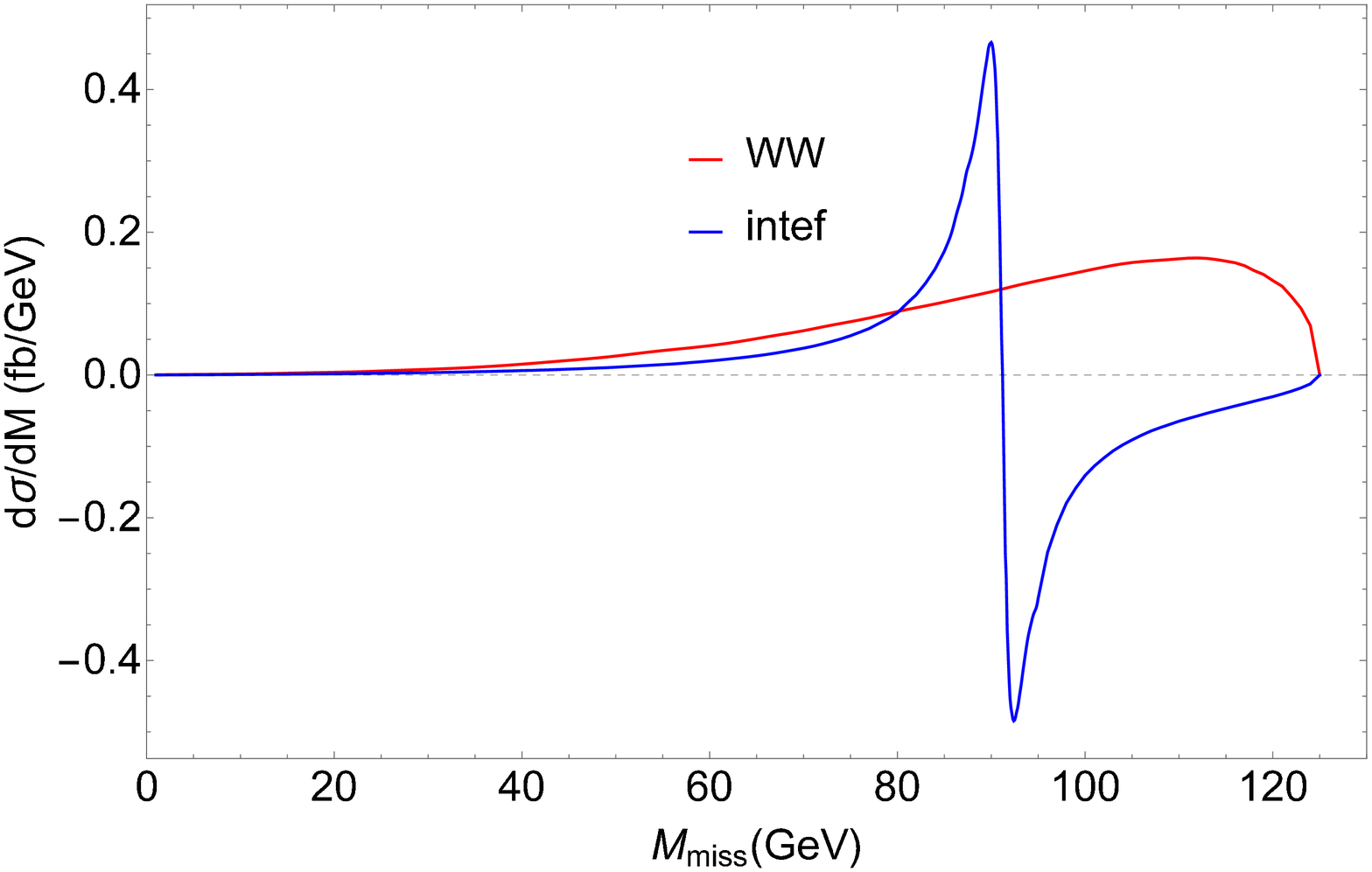}&
  \includegraphics[width=0.5\textwidth,clip]{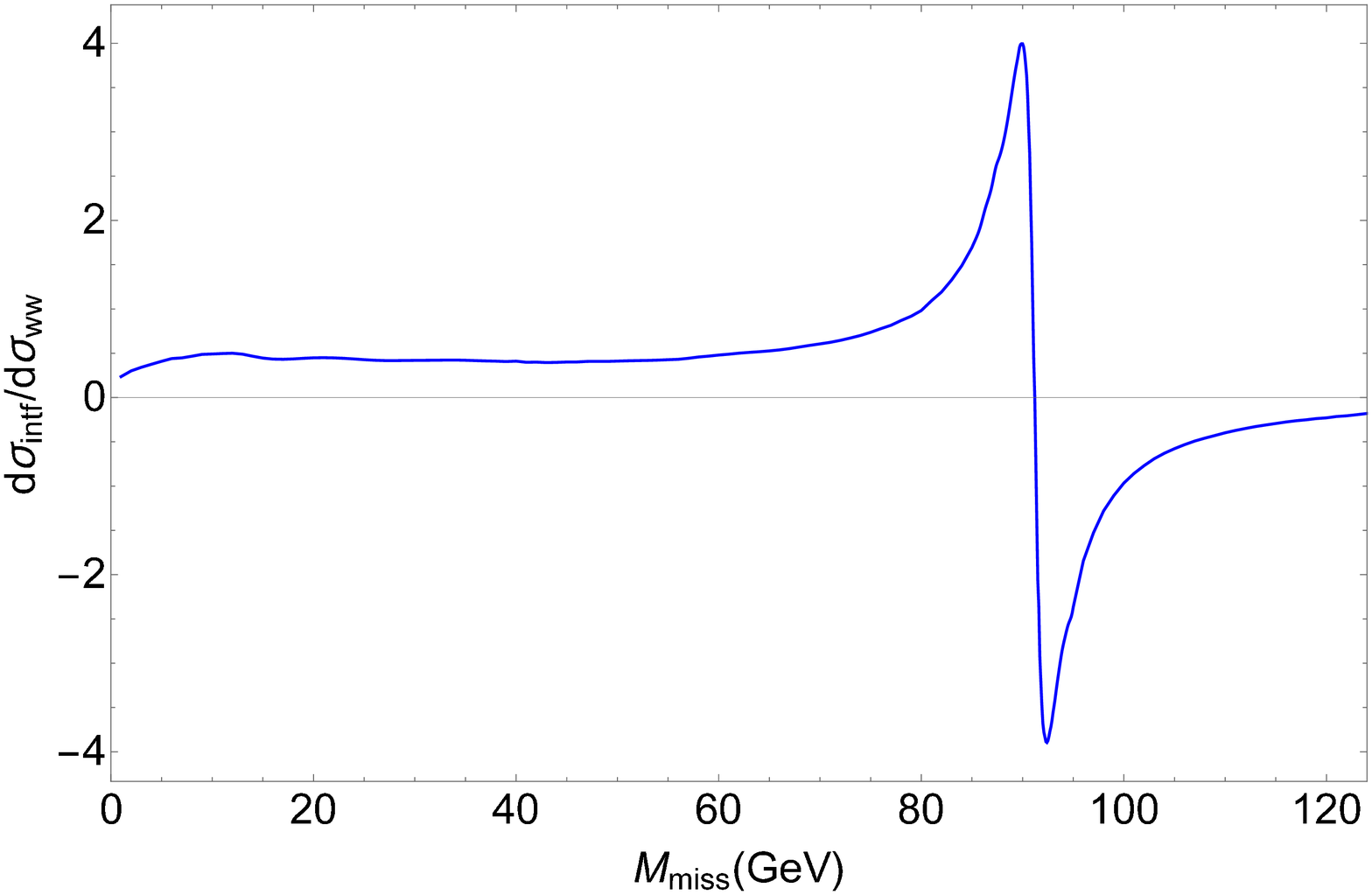}\\
  (a)&(b)\\
  \end{tabular}
    \caption{
    (a) Missing mass spectrums of $WW$ fusion term (red line) and the interference term (blue line);
    (b) the ratio between the missing mass distributions of the interference term and that of $WW$ fusion term $\ud\sigma_{\rm intf}/\ud\sigma_{WW}$ for different $M_{\rm miss}$.
    }
  \label{fig:dsigmamrecoil250}
\end{figure}
\begin{figure}
  \centering
  \begin{tabular}{cc}
  \includegraphics[width=0.5\textwidth,clip]{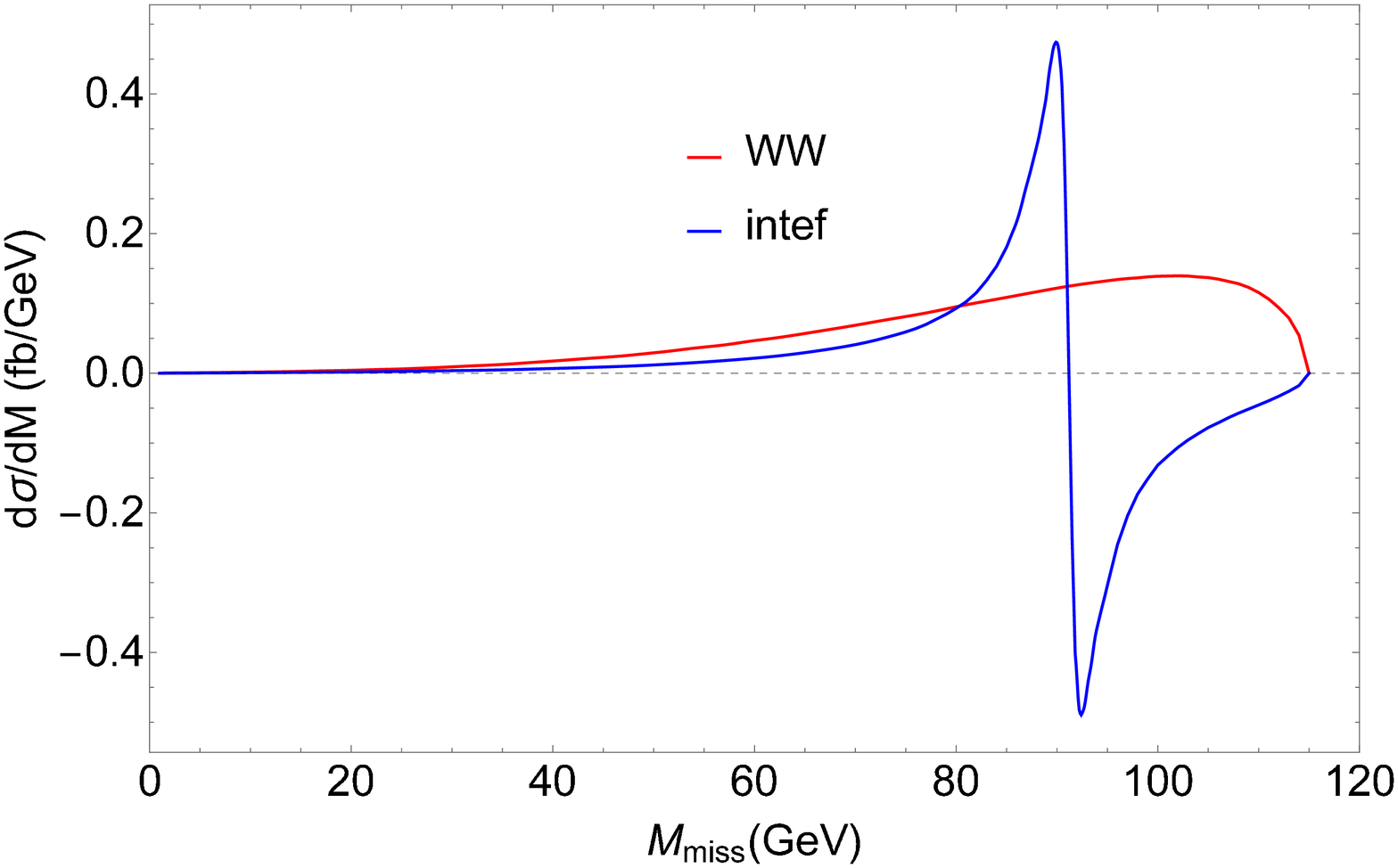}&
  \includegraphics[width=0.5\textwidth,clip]{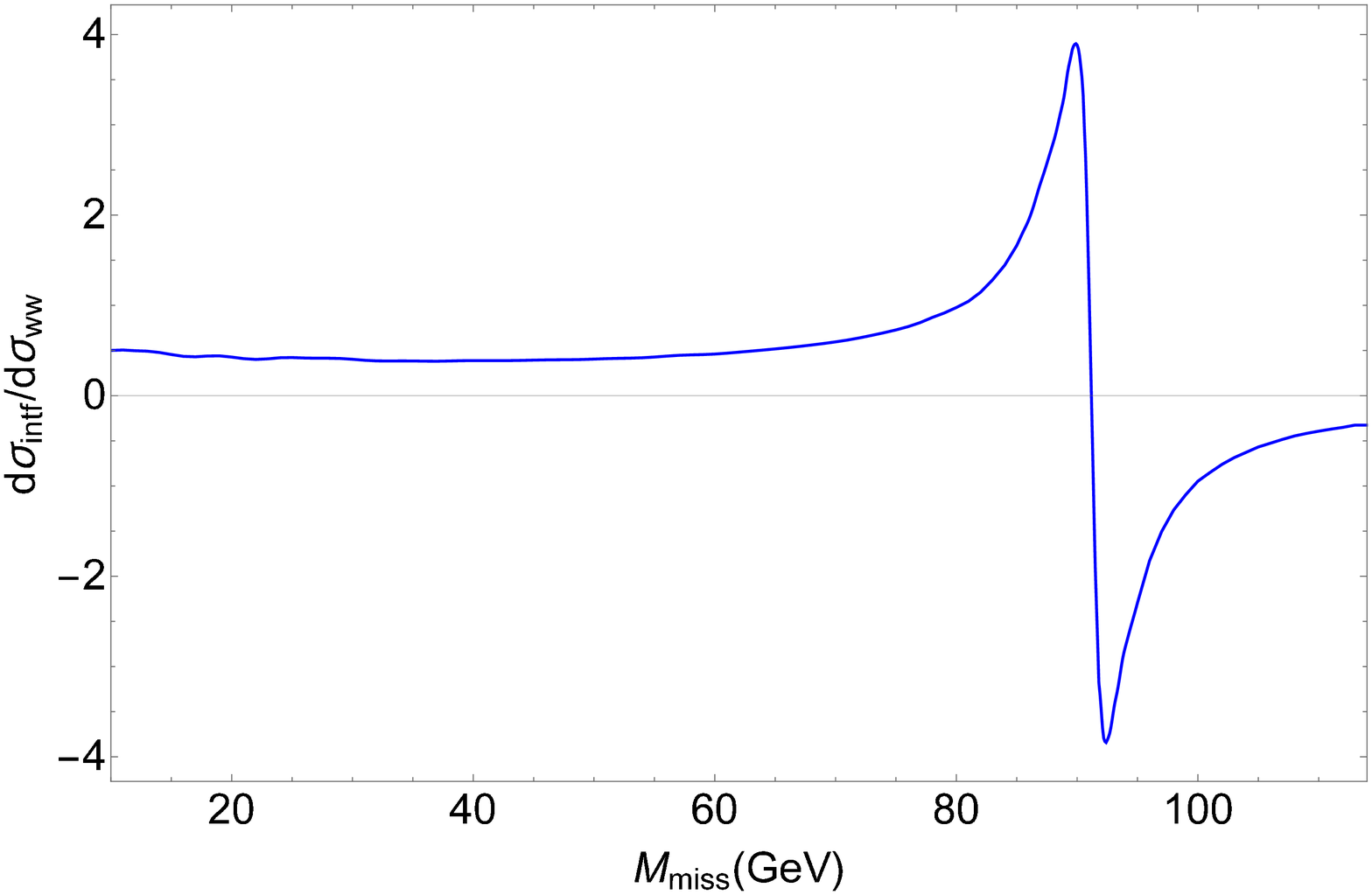}\\
  (a)&(b)\\
  \end{tabular}
    \caption{
    (a) Missing mass spectrums and (b) the ratio of the interference term to  $WW$ fusion term, same as Fig.~\ref{fig:dsigmamrecoil250} only at a different c.o.m energy $\sqrt{s}=240\gev$. The cross sections at $\sqrt{s}=240\gev$ are respectively $\sigma_{\rm intf}=0.444\fb$ and $\sigma_{WW}=6.36\fb$. Here the net effect of the interference term is positive.
    }
  \label{fig:dsigmamrecoil240}
\end{figure}
In this section, we display some differential distributions of both the signal and the background channels. Fig.~\ref{fig:dsigmacosh} and \ref{fig:dsigmamrecoil250} are respectively the angular distributions and recoil mass distributions for the $WW$ fusion process and the interference term at a c.m. energy $\sqrt{s}=250\gev$. Fig.~\ref{fig:dsigmamrecoil240} shows recoil mass distributions at a different c.m. energy $\sqrt{s}=240\gev$. The angular distribution of the interference term is uniformly negative and almost isotropic compared to that of the $WW$ fusion process, which peaks at $\theta_h\rightarrow 0$ and $\pi$. 
The recoil mass distributions of $WW$ fusion and the interference term are of comparable size as shown in Fig.~\ref{fig:dsigmamrecoil250} and \ref{fig:dsigmamrecoil240}. The differential cross section of $WW$ fusion process is more pronounced with higher recoil masses, while the interference term has two extreme values near the $Z$ pole. However, in conformity to the analytical results of scattering amplitudes in the last section, the interference term flips sign at the $Z$ pole. The two sides nearly cancel out and leave a negative net effect a few percents ($-4.1\%$) of $WW$ fusion process for $\sqrt{s}=250\gev$. While at $\sqrt{s}=240\gev$, the net effect of the interference term is positive.

\section{Impact on the Higgs physics}\label{sec4}

\begin{figure}
  \centering
		\includegraphics[scale=1,clip]{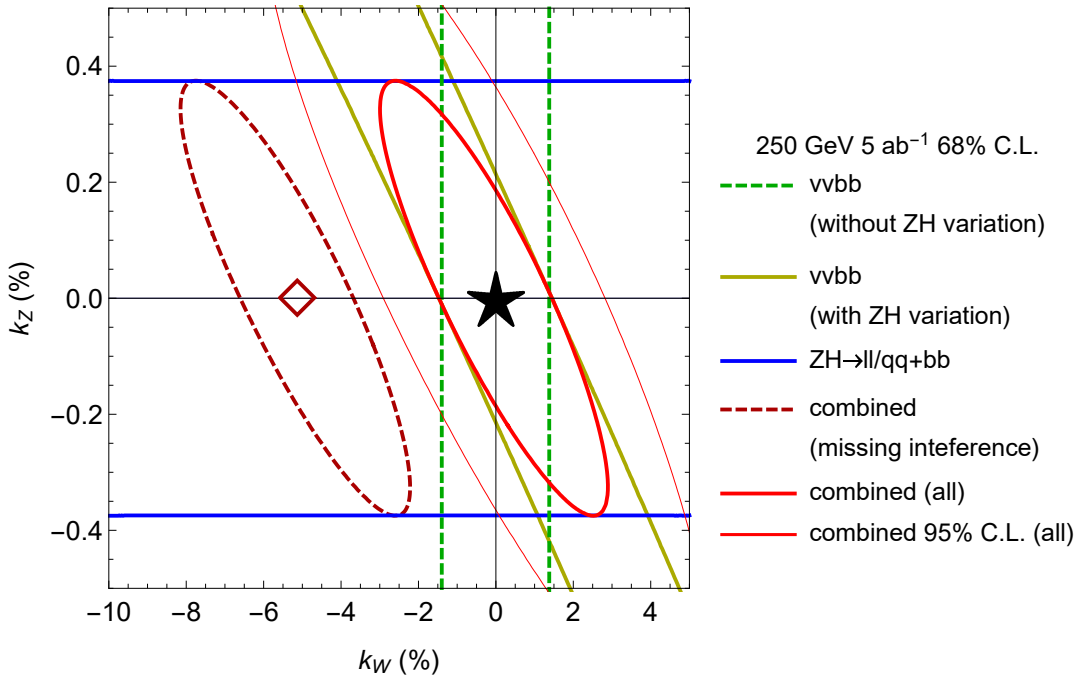}
		\caption{
			The interference effects reflected in the coupling extraction process in the $\kappa$-framework.
		}
		\label{fig:fit}
\end{figure}

We show the impact of the double interference effects in Fig.~\ref{fig:fit} in the $\kappa$-framework. First, in many lepton collider studies, the $\nu\bar\nu b\bar b$ final state are treated as a pure signal from $WW$-fusion with a SM background without theoretical uncertainty. However, as the dominant background is from $ZH\to \nu\bar\nu b\bar b$, which is also varying the fitting procedure, it shall not be treated as the background. In this figure, we show in green dashed green curves the $\Delta\chi^2=1$ window on $\kappa_W$ with the incorrect treatment of neglecting the variation from the $HZZ$ coupling. In contrast, we show in yellow curves the correct two-dimensional $\Delta\chi^2=1$ window in the $\kappa_Z$ and $\kappa_W$ plane. As one can see, a flat direction is caused from this consideration for the $\nu\nu h$ measurement. Furthermore, we can take advantage of the $ZH\to \ell\ell b\bar b$ and $ZH\to q\bar q b\bar b$ precision cross section measurement to constraint the $\kappa_Z$ direction, shown in the window bounded by blue curves. We note here one should not take the face value of the combined $ZH, H\to b\bar b$ cross section precision, as it has a dominant contribution actually from the $Z\to \nu\bar \nu $ process to avoid double counting.

Combining these measurements one can determine a 68\% C.L. precision on the $\kappa_W$ of 2.8\% instead of 1.4\% with the incorrect treatment of neglecting the interplay between $\kappa_W$ and $\kappa_Z$ for this process.

Next, the quantum interference effect between the $ZH\to \nu\bar \nu b\bar b$ amplitude and the $WW$-fusion with $H\to b\bar b$ amplitude yields roughly a 10\% reduction for the $WW$-fusion cross section. Hence, neglecting this effect will result in a wrong expectation value of the measurement, shown as the dark red square in the figure, which is 4-5$\sigma$ away from the true SM expectation!

\end{document}